\documentclass[twocolumn] {revtex4}
\usepackage{graphicx}
\begin{document}
\title{Two-mode excited entangled coherent states and their entanglement properties}
\author{Dong-Lin Zhou  and Le-Man Kuang \footnote{Corresponding author. Email: lmkuang@hunnu.edu.cn}}
\address{Laboratory of Low-Dimensional Quantum Structures and Quantum
Control of Ministry of Education, and Department of Physics, Hunan
Normal University, Changsha 410081, People's Republic of China}

\begin{abstract}
We introduce two types of two-mode excited entangled coherent
states (TMEECSs) $|\Psi_{\pm}(\alpha,m,n)\rangle$, study their
entanglement characteristics, and investigate the influence of
photon excitations on quantum entanglement. It is shown that for
the state $|\Psi_{+}(\alpha,m,n)\rangle$ the two-mode photon
excitations affect seriously entanglement character while the the
state $|\Psi_{-}(\alpha,m,n)\rangle$ is always a maximally
entangled state. We show how such states can be produced by using
cavity QED and quantum measurements. It is found that the
entanglement amount of the TMEECSs is larger than that of the
single-mode excited entangled coherent states with the same photon
excitation number.
\end{abstract}


\maketitle

\section{Introduction}

As it is well known that quantum entanglement has been viewed as
an essential resource for quantum information processing, and
creating and manipulating of entangled states are essential for
quantum information applications. Among these applications are
quantum computation \cite{nie}, quantum teleportation, and quantum
cryptography. In recent years, much attention has been paid to
continuous variable quantum information processing
\cite{cer,su,che,zha,brau,jeo,enk,ban,bra,fur,kz,zhou1,zhou2,zhou3,ck,gk,zk,kzk,mun}
in which continuous-variable-type entangled pure states play a key
role. For instance, two-state entangled coherent states (ECS) are
used to realize efficient quantum computation \cite{jeo} and
quantum teleportation \cite{enk}. Two-mode squeezed vacuum states
have been applied to quantum dense coding \cite{ban}. In
particular, following the theoretical proposal of Ref. \cite{bra},
continuous variable teleportation has been experimentally
demonstrated  for coherent states of a light field \cite{fur} by
using entangled two-mode squeezed vacuum states. Therefore, it is
an interesting topic to create and apply continuous-variable-type
entangled pure states. On the other hand, a coherent state is the
simplest continuous-variable state. Based on coherent states,  two
types of continuous-variable states, called  photon-added coherent
states \cite{aga} and entangled coherent states \cite{san},  have
been introduced and shown to have wide applications in both
quantum physics \cite{zav} and quantum information processing
\cite{jeo,enk,mun,wan}. In a previous paper \cite{xu}, single-mode
excited entangled coherent states (SMEECSs) are introduced. It has
been shown that the SMEECSs form a type of cyclic representation
of the Heisenberg-Weyl algebra and exhibit rich entanglement
properties. The purpose of this paper is to propose the concept of
two-mode excited entangled coherent states (TMEECSs), study their
preparation and entanglement properties.  This paper is organized
as follows. In Sec. II, we present the definition of the TMEECSs
and discuss their preparation. In Sec. III, we study entanglement
character of the TMEECSs and compare them with the SMEECSs. We
shall conclude this paper with discussions and remarks in the last
section.

\section{Two-mode excited entangled coherent states and their preparation}

In this section we introduce the definition of the TMEECSs and
present a possible scheme of producing them from the SMEECSs
through atom-field interaction. Let us begin with the following
two-mode entangled coherent states (ECSs)
\begin{equation}
\label{1}
|\Psi_{\pm}(\alpha,0)\rangle=\mathcal{A}_{\pm}(\alpha,0)(|\alpha,
\alpha\rangle  \pm |-\alpha, -\alpha\rangle ),
\end{equation}
where $|\alpha, \alpha\rangle=|\alpha\rangle\otimes|\alpha\rangle$
with $|\alpha\rangle=D(\alpha)|0\rangle$ being the usual Glauber
coherent state defined by the action of the displacement operator
$D(\alpha)=\exp(\alpha\hat{a}^{\dagger}-\alpha^*\hat{a})$ upon the
vacuum state $|0\rangle$. The normalization constants are given by
\begin{eqnarray}
\label{2} \mathcal{A}^{-2}_{\pm}(\alpha,0)=2\left[1 \pm \exp
(-4|\alpha|^{2})\right].
\end{eqnarray}

For convenience, we denote the first and second modes in two-mode
ECSs given by  Eq. (\ref{1}) by modes $a$ and $b$, respectively.
Then for the ECSs we consider $m$-photon excitations in  mode $a$
and $n$-photon excitations in  mode $a$  with $m\neq 0$ and $n\neq
0$, respectively, and introduce the TMEECSs defined by
\begin{equation}
\label{3} |\Psi_{\pm}(\alpha,m,n)\rangle=\mathcal{N}_{\pm}(\alpha,
m,n)\hat{a}^{\dagger m}\hat{b}^{\dagger m}(|\alpha, \alpha\rangle
\pm |-\alpha, -\alpha\rangle ),
\end{equation}
where  $\hat{a}^{\dagger}$ and  $\hat{b}^{\dagger}$ are the
creation operators of the modes $a$ and $b$, respectively. It is
straight forward to calculate the normalization constants in Eq.
(\ref{3}) with the following expressions
\begin{eqnarray}
\label{4}
\mathcal{N}^{-2}_{\pm}(\alpha,m,n)&=&2m!n!\left[L_m(-|\alpha|^2)L_n(-|\alpha|^2)\right.\nonumber\\
& &\left. \pm e^{-4|\alpha|^{2}}L_m(|\alpha|^2)L_n(|\alpha|^2)
\right],
\end{eqnarray}
where $L_m(x)$ is a Laguerre polynomials of order $m$ defined by
\begin{equation}
\label{5} L_m(x)=\sum^{m}_{n=0}\frac{(-1)^nm!x^n}{(n!)^2(m-n)!}.
\end{equation}
In the derivation of the Eq. (\ref{4}) we have used the following
expressions
\begin{eqnarray}
\label{6} \langle\alpha|\hat{a}^{m}\hat{a}^{\dagger
m}|\alpha\rangle&=&m!L_m(-|\alpha|^2), \nonumber\\
\langle\alpha|\hat{a}^{m}\hat{a}^{\dagger
m}|-\alpha\rangle&=&m!e^{-2|\alpha|^{2}}L_m(|\alpha|^2),
\hspace{0.3cm}(m\neq 0).
\end{eqnarray}

If we introduce the following normalized states with respect to
mode $a$ and  mode $b$
\begin{eqnarray}
\label{7} |a(\pm\alpha,m)\rangle&=&N(\alpha,m)\hat{a}^{\dagger
m}|\pm\alpha\rangle,\nonumber\\
|b(\pm\alpha,m)\rangle&=&N(\alpha,m)\hat{b}^{\dagger
m}|\pm\alpha\rangle,
\end{eqnarray}
where the normalized coefficients are given by
\begin{equation}
\label{8} N^{-2}(\alpha,m)=m!L_m(-|\alpha|^2).
\end{equation}

Then the TMEECSs $|\Psi_{\pm}(\alpha,m,n)\rangle$ in terms of four
normalized states can be rewritten as
\begin{eqnarray}
\label{9}
|\Psi_{\pm}(\alpha,m)\rangle&=&\mathcal{M}_{\pm}(\alpha,m,n)[
|a(\alpha,m)\rangle\otimes|b(\alpha,n)\rangle \nonumber\\
&&\pm |a(-\alpha,m)\rangle\otimes|b(-\alpha,n)\rangle],
\end{eqnarray}
where the normalization constants are given by
\begin{widetext}
\begin{equation}
\label{10}
 \mathcal{M}^2_{\pm}(\alpha,m,n)=\frac{L_m(-|\alpha|^2)L_n(-|\alpha|^2)}{2\left[L_m(-|\alpha|^2)L_n(-|\alpha|^2)
\pm e^{-4|\alpha|^{2}}L_m(|\alpha|^2)L_n(|\alpha|^2)  \right]}.
\end{equation}
\end{widetext}

We now present a possible scheme to produce them from the two mode
ECSs through atom-field interaction. Consider an interaction
between a two-level atom with a cavity field. The atom makes a
transition from the excited state $|e\rangle$ to the ground state
$|g\rangle$ by emitting a photon. In the interaction picture, the
Hamiltonian of resonant interaction is given by
\begin{equation}
\label{11} \hat{H}=\hat{\sigma}_{+}\hat{b} +
g^*\hat{\sigma}_{-}\hat{b}^{\dagger},
\end{equation}
where $\hat{\sigma}_{+}$ and $\hat{\sigma}_{-}$ are Pauli operator
corresponding to the two-level atom, $\hat{b}^{\dagger}$ and
$\hat{b}$ are the creation and annihilation operator of the field
mode $b$, $g$ is the coupling constant.

Suppose that the atom is initially in the excited state and the
two field modes is initially in the SMEECS with $m$ photon
excitations ($m\neq 0$) given by
\begin{equation}
\label{12} |\Psi_{\pm}(\alpha,m)\rangle=N_{\pm}(\alpha,
m)\hat{a}^{\dagger m}(|\alpha, \alpha\rangle  \pm |-\alpha,
-\alpha\rangle ),
\end{equation}
where the normalization constant is given by
\begin{eqnarray}
\label{13} N^{-2}_{\pm}(\alpha,m)=2m!\left[L_m(-|\alpha|^2) \pm
e^{-2|\alpha|^{2}}L_m(|\alpha|^2)  \right],
\end{eqnarray}
where $L_m(x)$ is $m$-th Laguerre polynomial defined in Eq.
(\ref{5}).

Suppose that only the mode $b$ interacts with the atom. Then for
the weak coupling case, the state of the atom-field system at time
$t$ can be approximated by
\begin{equation}
\label{14} |\psi(t)\rangle \approx |\Psi_{\pm}(\alpha,m)\rangle
\otimes|e\rangle
-i\hat{H}t|\Psi_{\pm}(\alpha,m)\rangle\otimes|e\rangle,
\end{equation}
which is approximately valid for interaction times such that
$gt\ll 1$. Making use of Eq. (\ref{11}), one can reduce (\ref{14})
to
\begin{equation}
 \label{15}
|\psi(t)\rangle \approx
|\Psi_{\pm}(\alpha,m)\rangle\otimes|e\rangle
-i(g^*t)\hat{b}^{\dagger}|\Psi_{\pm}(\alpha,m)\rangle\otimes|g\rangle,
\end{equation}
which indicates that if the atom is detected to be in the ground
state $|g\rangle$, then after normalization the state of the two
optical fields is reduced to the TMEECS
$|\Psi_{\pm}(\alpha,m,1)\rangle$ given by Eq. (\ref{3}). If we
consider a succession of $m$ atoms through the cavity and if we
detect all the atoms in the ground state $|g\rangle$, then the
state of the two optical fields is reduced to the desired state,
the TMEECS with $(m+n)$-photon excitations. Hence, we can, in
principle, produce the TMEECS  $|\Psi_{\pm}(\alpha, m,n)\rangle$.

\section{The entanglement amount of  TMEECS}

In this section, we calculate the amount of entanglement of the
TMEECSs and investigate the influence of the photon excitations on
the entanglement of the TMEECSs. From Eq. (\ref{9}) we can see
that the TMEECSs are two-component entangled states. The degree of
quantum entanglement of the two-state entangled states can be
measured in terms of the concurrence \cite{kz,wan,hill} which is
generally defined for discrete-variable entangled states to be
\cite{hill}
\begin{equation}
\label{16}
\mathcal{C}=|\langle\Psi|\sigma_y\otimes\sigma_y|\Psi^*\rangle|,
\end{equation}
where $|\Psi^*\rangle$ is the complex conjugate of $|\Psi\rangle$.
The concurrence equals one for a maximally entangled state.

For continuous-variables-type entangled states like (\ref{9}), we
consider a general bipartite entangled state
\begin{equation}
\label{17} |\psi\rangle=\mu|\eta\rangle\otimes|\gamma\rangle +
\nu|\xi\rangle\otimes|\delta\rangle,
\end{equation}
where $|\eta\rangle$ and $|\xi\rangle$ are {\it normalized} states
of subsystem $1$ and  $|\gamma\rangle$ and $|\delta\rangle$ {\it
normalized} states of subsystem $2$ with complex $\mu$ and $\nu$.
Through transforming continuous-variables-type components to
discrete orthogonal basis and making use of a Schmidt
decomposition \cite{man}, it is found the concurrence of the
entangled state (\ref{17}) to be given by the following expression
\cite{kz,run}
\begin{eqnarray}
\label{18}
\mathcal{C}&=&\frac{2|\mu||\nu|\sqrt{(1-|p_1|^2)(1-|p_2|^2)}}{|\mu|^2+|\nu|^2
+ 2 \tt{Re} (\mu^*\nu p_1p^*_2)},
\end{eqnarray}
where the two overlapping functions are defined by
\begin{equation}
\label{19} p_1=\langle\eta|\xi\rangle, \hspace{0.3cm}
p_2=\langle\delta|\gamma\rangle.
\end{equation}

For the case of the no-photon excitation ECSs defined in Eq.
(\ref{1}), making use of Eqs. (\ref{1}) and (\ref{2}) from Eqs.
(\ref{18})-(\ref{19}) we can find that the concurrence is given by
\begin{equation}
\label{20} \mathcal{C}_{-}(\alpha, 0)=1, \hspace{0.3cm}
\mathcal{C}_{+}(\alpha,
0)=\frac{1-e^{-4|\alpha|^2}}{1+e^{-4|\alpha|^2}},
\end{equation}
which implies that the degree of entanglement of the ECS
$|\Psi_{-}(\alpha,0)\rangle$ is independent of the state parameter
$\alpha$, and it is a maximally entangled state while the amount
of entanglement of the ECS $|\Psi_{+}(\alpha,0)\rangle$ is less
than that of the ECS $|\Psi_{-}(\alpha,0)\rangle$. The concurrence
$\mathcal{C}_{+}(\alpha, 0)$ increases with $|\alpha|^2$, and the
state $|\Psi_{+}(\alpha,0)\rangle$ approaches the maximally
entangled coherent state with $\mathcal{C}_{+}(\alpha, 0)\approx
1$ for the strong filed case of the large $|\alpha|^2$.

When there exist photon excitations for both modes, i.e., $m \neq
0$ and $n \neq 0$, the TMEECSs are given by Eqs. (\ref{3}) and
(\ref{9}). From Eqs. (\ref{18}) and (\ref{19}) we find the
corresponding concurrence to be
\begin{equation}
\label{21} \mathcal{C}_{\pm}(\alpha,
m,n)=\frac{\sqrt{(1-|p_1(\alpha, m)|^2)(1-|p_2(\alpha,
n)|^2)}}{1\pm p_1(\alpha, m)p_2(\alpha,n)},
\end{equation}
where the two overlapping functions are given by
\begin{eqnarray}
\label{22} p_1(\alpha,
m)&=&e^{-2|\alpha|^{2}}\frac{L_{m}(|\alpha|^2)}{L_{m}(-|\alpha|^2)},
\nonumber\\
p_2(\alpha,
n)&=&e^{-2|\alpha|^{2}}\frac{L_{n}(|\alpha|^2)}{L_{n}(-|\alpha|^2)}.
\end{eqnarray}

From  Eqs. (\ref{21}-\ref{22}) we can see that the TMEECSs exhibit
the exchanging symmetry with respect to the exchange of two-mode
photon excitations
\begin{eqnarray}
\label{23} \mathcal{C}_{\pm}(\alpha,
m,n)&=&\mathcal{C}_{\pm}(\alpha,n,m),
\nonumber\\
\mathcal{C}_{\pm}(\alpha,
m+k,n)&=&\mathcal{C}_{\pm}(\alpha,m,n+k),
\end{eqnarray}
where $k$ is an arbitrary non-negative integer. Eq. (\ref{23})
indicates that the allotment of photon excitations between two
modes does not affect the entanglement amount of the TMEECSs when
the total number of photon excitations are fixed.

In order to observe the influence of the photon excitations on the
quantum entanglement of the TMEECS
$|\Psi_{\pm}(\alpha,m,n)\rangle$, we consider the case of $m=n$ in
which there are the same photon excitations in each filed modes of
the TMEECS.

Firstly, let us consider the TMEECS
$|\Psi_{+}(\alpha,m,m)\rangle$. In this case we find the
corresponding concurrence to be
\begin{eqnarray}
\label{24} \mathcal{C}_{+}(\alpha, m,m)&=&\frac{1-p^2_1(\alpha,
m)}{1+p^2_1(\alpha, m)},
\end{eqnarray}
which indicates that the amount of entanglement of the TMEECS
$|\Psi_{\pm}(\alpha,m,m)\rangle$ decreases with the increase of
the overlapping function $|p_1(\alpha, m)|$. In the weak field
regime of $|\alpha|^2\ll 1$, we have $L_{m}(|\alpha|^2)\approx
1-m|\alpha|^2$. Then from  Eqs. (\ref{22}) and (\ref{24}) we can
get
\begin{eqnarray}
\label{25} \mathcal{C}_{+}(\alpha, m,m)&\approx&(1+m)|\alpha|^2,
\end{eqnarray}
which implies that the concurrence increases with the increase of
the two-mode photon excitations. Hence in the weak field regime
the photon excitation can enhance the entanglement amount for the
TMEECS $|\Psi_{+}(\alpha,m,m)\rangle$.

It would be interesting to compare entanglement character of the
TMEECS $|\Psi_{+}(\alpha,m,m)\rangle$ which contains $2m$ photon
excitations that of the SMEECS $|\Psi_{+}(\alpha,2m)\rangle$ which
involves $2m$ photon excitations as well. The SMEECSs
$|\Psi_{\pm}(\alpha,2m)\rangle$ with $m\neq 0$ \cite{xu} are
defined by
\begin{equation}
\label{26} |\Psi_{\pm}(\alpha,2m)\rangle=\mathcal{N}_{\pm}(\alpha,
2m)\hat{a}^{\dagger 2m}(|\alpha, \alpha\rangle  \pm |-\alpha,
-\alpha\rangle ),
\end{equation}
where the normalization constants Eq. (\ref{3}) are given by
\begin{equation}
\label{27}
\mathcal{N}^{-2}_{\pm}(\alpha,2m)=4m!\left[L_{2m}(-|\alpha|^2) \pm
e^{-4|\alpha|^{2}}L_{2m}(|\alpha|^2)  \right].
\end{equation}

The concurrence of the SMEECS $|\Psi_{+}(\alpha,2m)\rangle$ is
given by the following expression
\begin{widetext}
\begin{equation}
\label{28}
 \mathcal{C}_{+}(\alpha, 2m)=\frac{
\left[\left(1-e^{-4|\alpha|^2}\right)\left(L^2_{2m}(-|\alpha|^2)-e^{-4|\alpha|^{2}}L^2_{2m}(|\alpha|^2)\right)\right]^{1/2}}
{L^2_{2m}(-|\alpha|^2)+e^{-4|\alpha|^2}L^2_{2m}(|\alpha|^2)},
\end{equation}
\end{widetext}
which implies that in the weak field regime of
$|\alpha|^2\ll 1$, we can get
\begin{eqnarray}
\label{29}
 \mathcal{C}_{+}(\alpha, 2m)&=&\sqrt{1+2m}|\alpha|^2,
\end{eqnarray}
which indicates that the photon excitation can enhance the
entanglement amount for the SMEECS $|\Psi_{+}(\alpha,m)\rangle$.
In particular, when $|\alpha|^2\ll 1$, we have
$\mathcal{C}_{+}(\alpha, 2m)\ll 1$. Therefore, in the weak field
regime the TMEECS $|\Psi_{\pm}(\alpha,m,m)\rangle$ exhibits
similar entanglement character to that of the SMEECS
$|\Psi_{+}(\alpha,m)\rangle$.

Secondly, we consider the TMEECS $|\Psi_{-}(\alpha,m,m)\rangle$.
In this case the corresponding concurrence is found to be
\begin{eqnarray}
\label{30} \mathcal{C}_{-}(\alpha, m,m)&=&1,
\end{eqnarray}
which implies that the TMEECS $|\Psi_{-}(\alpha,m,m)\rangle$ is
always a maximally entangled state and $2m$ photon excitations do
not affect the entanglement amount of the state. This property is
very different from that of the SMEECS
$|\Psi_{-}(\alpha,2m)\rangle$ defined in Eq. (\ref{26}). The
concurrence of the SMEECS $|\Psi_{-}(\alpha,2m)\rangle$ is given
by
\begin{widetext}
\begin{equation}
\label{31} \mathcal{C}_{-}(\alpha, 2m)=\frac{
\left[\left(1-e^{-4|\alpha|^2}\right)\left(L^2_{2m}(-|\alpha|^2)-e^{-4|\alpha|^2}L^2_{2m}(|\alpha|^2)\right)\right]^{1/2}}
{L^2_{2m}(-|\alpha|^2)-e^{-4|\alpha|^2}L^2_{2m}(|\alpha|^2)}.
\end{equation}
\end{widetext}
In the weak field regime of $|\alpha|^2\ll 1$, we
find that
\begin{eqnarray}
\label{32} \mathcal{C}_{-}(\alpha,
2m)&\approx&\frac{1}{\sqrt{1+2m}},
\end{eqnarray}
which indicates that the concurrence $\mathcal{C}_{-}(\alpha, 2m)$
decreases with the increase of the photon excitation number $2m$.
In particular, when $m\gg 1$ we have
$\mathcal{C}_{-}(\alpha,2m)\ll 1$. Hence, the photon excitation
suppresses the amount of entanglement for the the SMEECS
$|\Psi_{-}(\alpha,2m)\rangle$ in the weak field regime.

\section{Concluding Remarks}

We have proposed two types of TMEECSs
$|\Psi_{\pm}(\alpha,m,n)\rangle$, studied their entanglement
characteristics, and investigated the influence of photon
excitations on quantum entanglement. We have indicated it is
possible to produce such states by using cavity QED and quantum
measurements. It is found that the two TMEECSs
$|\Psi_{\pm}(\alpha,m,n)\rangle$ exhibit quite different
entanglement properties. In particular, for the state
$|\Psi_{+}(\alpha,m,m)\rangle$ the two-mode photon excitations
affect seriously entanglement character, and the entanglement
amount decreases with the two-mode photon excitations in the weak
field regime.  However, the state $|\Psi_{-}(\alpha,m,m)\rangle$
is always a maximally entangled state, the two-mode photon
excitations does not change the amount entanglement of the state.
We have also made comparisons between the TMEECSs
$|\Psi_{\pm}(\alpha,m,m)\rangle$ and the SMEECSs
$|\Psi_{\pm}(\alpha,2m)\rangle$. It has been shown that two-mode
photon excitations have more advantages than single-mode photon
excitations such as the entanglement amount of the TMEECSs
$|\Psi_{\pm}(\alpha,m,m)\rangle$ is larger than that of the
SMEECSs $|\Psi_{\pm}(\alpha,2m)\rangle$ for the same photon
excitation number $2m$. This approach of enhancing entanglement by
using two-mode excitations of continuous-variable quantum sates
opens a new way to create new entanglement resources with
continuous variables.

\acknowledgments This work was supported by the National
Fundamental Research Program Grant No.  2007CB925204, the National
Natural Science Foundation under Grant Nos. 10775048 and 10325523,
the Foundation of the Education Ministry of China, and the
Education Committee of Hunan Provinceunder Grant No. 08W012.

\end{document}